\documentclass[12pt]{article}

\usepackage[letterpaper,hmargin=1in,vmargin=1in]{geometry}

\usepackage{graphicx,epstopdf,amsmath,amsfonts,amssymb}

\def\be{\begin{equation}}
\def\ee{\end{equation}}
\def\ba{\begin{eqnarray}}
\def\ea{\end{eqnarray}}
\def\ge{\mathrel{\raise.3ex\hbox{$>$\kern-.75em\lower1ex\hbox{$\sim$}}}}
\def\la{\mathrel{\raise.3ex\hbox{$<$\kern-.75em\lower1ex\hbox{$\sim$}}}}

\def\simgt{\mathrel{\raise.3ex\hbox{$>$\kern-.75em\lower1ex\hbox{$\sim$}}}}
\def\simlt{\mathrel{\raise.3ex\hbox{$<$\kern-.75em\lower1ex\hbox{$\sim$}}}}

\newcommand{\bi}[1]{\bibitem{#1}}
\newcommand{\fr}[2]{\frac{#1}{#2}}

\newcommand{\nc}{\newcommand}

\nc{\gone}{\bar g_{\pi NN}^{(1)}}
\nc{\gzero}{\bar g_{\pi NN}^{(0)}}
\nc{\al}{\alpha}
\nc{\ga}{\gamma}
\nc{\de}{\delta}
\nc{\ep}{\epsilon}
\nc{\ze}{\zeta}
\nc{\et}{\eta}
\nc{\ka}{\kappa}
\nc{\rh}{\rho}
\nc{\si}{\sigma}
\nc{\ta}{\tau}
\nc{\up}{\upsilon}
\nc{\ph}{\phi}
\nc{\ch}{\chi}
\nc{\ps}{\psi}
\nc{\om}{\omega}
\nc{\Ga}{\Gamma}
\nc{\De}{\Delta}
\nc{\La}{\Lambda}
\nc{\Si}{\Sigma}
\nc{\Up}{\Upsilon}
\nc{\Ph}{\Phi}
\nc{\Ps}{\Psi}
\nc{\Om}{\Omega}
\nc{\ptl}{\partial}
\nc{\del}{\nabla}
\nc{\ov}{\overline}
\nc{\newcaption}[1]{\centerline{\parbox{15cm}{\caption{#1}}}}
\nc{\us}{U(1)$_S$}

\def\beq{\begin{equation}}
\def\eeq{\end{equation}}
\def\bmat{\begin{displaymath}}
\def\emat{\end{displaymath}}
\def\bear{\begin{eqnarray}}
\def\eear{\end{eqnarray}}
\def\ba{\begin{eqnarray}}
\def\ea{\end{eqnarray}}
\def\bery{\begin{array}}
\def\ery{\end{array}}
\def\bit{\begin{itemize}}
\def\eit{\end{itemize}}
\def\ben{\begin{enumerate}}
\def\een{\end{enumerate}}
\def\btab{\begin{tabular}}
\def\etab{\end{tabular}}
\def\btbl{\begin{table}}
\def\etbl{\end{table}}
\def\bfig{\begin{figure}[htb]}
\def\efig{\end{figure}}
\def\bpic{\begin{picture}}
\def\epic{\end{picture}}

\def\ga{\mathrel{\raise.3ex\hbox{$>$\kern-.75em\lower1ex\hbox{$\sim$}}}}
\def\la{\mathrel{\raise.3ex\hbox{$<$\kern-.75em\lower1ex\hbox{$\sim$}}}}
\def\gappeq{\mathrel{\rlap {\raise.5ex\hbox{$>$}}
{\lower.5ex\hbox{$\sim$}}}}
\def\lappeq{\mathrel{\rlap{\raise.5ex\hbox{$<$}}
{\lower.5ex\hbox{$\sim$}}}}

\def\gyr{{\rm \, G\kern-0.125em yr}}
\def\mev{{\rm \, Me\kern-0.125em V}}
\def\gev{{\rm \, Ge\kern-0.125em V}}
\def\tev{{\rm \, Te\kern-0.125em V}}

\begin{document}

\begin{titlepage}

\setcounter{page}{1}

\vspace*{0.2in}

\begin{center}

\hspace*{-0.6cm}\parbox{17.5cm}{\Large \bf \begin{center}

Probing a Secluded U(1) at \boldmath{$B$}-factories
\end{center}}

\vspace*{0.5cm}
\normalsize

\vspace*{0.5cm}
\normalsize

{\bf Brian Batell$^{\,(a)}$, Maxim Pospelov$^{\,(a,b)}$, and Adam Ritz$^{\,(b)}$}

\smallskip
\medskip

$^{\,(a)}${\it Perimeter Institute for Theoretical Physics, Waterloo,
ON, N2J 2W9, Canada}

$^{\,(b)}${\it Department of Physics and Astronomy, University of Victoria, \\
     Victoria, BC, V8P 1A1 Canada}

\smallskip
\end{center}
\vskip0.2in

\centerline{\large\bf Abstract}

A secluded U(1)$_S$ gauge field, kinetically mixed with Standard Model hypercharge, provides a
`portal' mediating interactions with a hidden sector at the renormalizable level, as recently exploited
in the context of WIMP dark matter. The U(1)$_S$ symmetry-breaking scale 
may naturally be suppressed relative to the weak scale, and so this sector is efficiently probed by medium 
energy $e^+e^-$ colliders. We study the collider signatures of the minimal U(1)$_S$ model, focusing on the reach 
of $B$-factory experiments such as BaBar and BELLE. In particular, we show that Higgs-strahlung 
 in the secluded sector can lead to multi-lepton signatures which probe the natural range for 
 the kinetic mixing angle $\kappa\sim 10^{-2}-10^{-3}$ over a large portion of the kinematically accessible
parameter space.

\vfil
\leftline{March 2009}
    
\end{titlepage}

\subsection*{1. Introduction}

The Standard Model (SM) allows relatively few interactions with new physics at the level of relevant or marginal operators. These `portals' - comprising specific couplings
to the Higgs sector, to right-handed neutrinos, and kinetic mixing with U(1) hypercharge - are thus of considerable 
interest from a general phenomenological perspective as the channels where new physics, and a hidden sector for example, may be probed with maximal sensitivity. Of these, it is 
clear that the latter portal, kinetic mixing of a new \us\ gauge field with hypercharge U(1)$_Y$, $\kappa F^S_{\mu\nu} F^Y_{\mu\nu}$ \cite{holdom}, stands out 
in terms of  current detection capabilities since it implies a renormalizable coupling with the photon and the $Z$, whose properties are of course extremely well-measured. 
This presence of a secluded \us\ sector, under which the SM degrees of freedom are uncharged, also admits a natural hypothesis about the scale of the
interaction and the mass of the \us\ vector  that we will denote as $V$. In particular, $\ka \sim 10^{-4} - 10^{-2}$, and $m_V \sim \ka~\times$~(weak scale)
are the relations that would naturally arise from radiative generation of kinetic mixing, and an associated radiative transfer of symmetry breaking from the SM to \us.

This idea has received a great deal of attention in recent months in the context of models of dark matter \cite{review}, with the Weakly Interacting Massive Particle (WIMP) 
dark matter candidate belonging to the secluded \us\ sector. In particular, the  secluded regime with $m_V < m_{\rm WIMP}$ can lead to dramatically different 
phenomenology as compared to the standard WIMP scenario, in terms of both direct and indirect detection \cite{PRV}. Perhaps the most intriguing aspect of 
these theories is that a relatively modest hierarchy of scales, $m_V \sim {\cal O}({\rm GeV}) \ll m_{\rm WIMP}$ 
naturally leads to an enhanced WIMP annihilation cross section in the galaxy \cite{AFSW,PR} and to the 
enhancement and/or dominance of the leptonic branching fraction in the annihilation products \cite{cholis}.  
Thus models of this type can naturally accommodate the recent observations of excess cosmic ray positrons by PAMELA \cite{pamela} and 
the total electron and positron flux by ATIC \cite{atic}, should these anomalies ultimately be related
to WIMP dark matter.

Through the kinetic mixing portal, a secluded \us\ sector at or below the GeV-scale will leave its imprints on low-energy phenomenology. 
Recently, Ref. \cite{tests} examined the constraints arising from precision QED tests as well as radiative decays of strange particles. These 
constraints are fairly model-independent, relying only on the mass of the \us\ vector and its admixture with the photon. However, the constraints 
deteriorate rather quickly for $ m_V $ above a few hundred MeV. Moreover, the main constraint coming from the measurement of the 
anomalous magnetic moment of the muon is difficult to implement, given the current discrepancy with SM predictions (see {\em e.g.} \cite{g-2}). 
On the other hand, because the new \us\ vectors couple only to vector currents, dedicated collider probes of the \us\ sector are also
a powerful tool and can potentially provide sensitivity in substantial regions of the $(\ka,m_V)$ parameter plane.

In this paper, we will investigate the signatures of a minimal secluded \us\ extension of the SM at medium energy $e^+e^-$ colliders. 
This minimal model assumes the existence of an elementary Higgs$'$ boson, which spontaneously breaks the U(1)$_S$ symmetry, but 
many features of the model will be replicated in more intricate realizations of the secluded gauge and Higgs$'$ sectors. Since the Higgs$'$ sector is necessarily
charged under \us\, the direct production cross-section scales as ${\cal O}(\kappa^2)$.  Medium energy $e^+e^-$ colliders with $\sqrt{s}\sim$ few GeV and high luminosity
are clearly an ideal tool to detect new particles with masses in the GeV range, as production cross sections fall as $1/s$. In particular, the $B$-factories 
BaBar and BELLE are well positioned to probe a secluded sector because of their large data sets, with each having an integrated luminosity of 
order $\sim 500\, {\rm fb}^{-1}$. The phenomenology will depend quite sensitively on assumptions about the particle content in the hidden sector, but as
the natural first step we will assume that any extra hidden-sector particles ({\em e.g.} the WIMP candidate) are heavier than the minimal set comprising the 
\us\ gauge boson and the Higgs$'$, which we assume to have ${\cal O}$(GeV) masses. The decay channels for these particles will thus have SM degrees of freedom 
in the final state. Because of the limited number of these channels, and suppression from the small kinetic mixing angle, the vector and Higgs$'$ are generally 
quite narrow. As a consequence,  $s$-channel resonant production of vectors is extremely unlikely, leading us to consider 
two-particle production channels.

We will focus on the Higgs$'$-strahlung process which is particularly interesting for a number of reasons. It is one of the few production processes with an 
amplitude suppressed by just a single power of the kinetic mixing angle and can therefore readily occur for $\ka\sim {\cal O}(10^{-2}-10^{-3})$. In many ways this 
production mechanism parallels conventional Higgs-strahlung in the SM. Although the vector will in general have a substantial branching to lepton pairs, 
the decays of the Higgs$'$ will depend on its mass relative to that of the vector. If the Higgs$'$ is heavy it will decay to two vectors, eventually leading to a six 
lepton final state. On the other hand, if the Higgs$'$ is light it will decay via loop processes to leptons and possibly hadrons, in which case it is long-lived 
and will most-likely appear as missing energy.

The collider phenomenology of additional light hidden U(1)$'$ bosons has been considered previously in the context of 
MeV-scale dark matter models \cite{mev,uboson1,uboson2}, where it was assumed that both dark matter (DM) and the SM are charged under  U(1)$'$, 
and there is a strong hierarchy of couplings, $g'_{SM} \ll g'_{DM}$. Such models are qualitatively different from the 
one we consider here, as in Refs.~\cite{mev,uboson1,uboson2} all mediators have a much larger 
invisible decay width to DM than to the SM.  Signatures of these models at $B$-factories, relevant to the scenario considered here, 
were studied in \cite{BCD}, though not the Higgs$'$-strahlung process. The hadron collider phenomenology of a non-minimal dark sector coupled through
kinetic mixing was first  surveyed in \cite{AW}, and 
has been elaborated on recently in \cite{nonabelian}. We will make some comments on 
the sensitivity of hadron colliders to the minimal scenario below. For other recent work on U(1)$'$ scenarios covering different parameter regimes
and phenomenology, see \cite{langacker,nw,prv2,jrr}.

The organization of this paper is as follows. In Sec.~2 we present the minimal secluded \us\ model with a single complex scalar serving as the Higgs$'$. 
We also provide details of the relevant decay channels and rates. We consider the $B$-factory collider signals in Sec.~3, focusing on the 
Higgs$'$-strahlung process. An overview of the backgrounds is given for different kinematical regions, and we present estimates of the optimal 
sensitivities that $B$-factories could achieve in probing the \us\ sector. We conclude in Sec.~4 with further discussion of a number of other signatures and 
possible directions to explore.

\subsection*{2. Secluded U(1)$_S$ and the Higgs$'$ sector}

We will consider the minimal implementation of a secluded U(1)$_S$. We add to the SM a \us\ gauge boson $V_\mu$ and a single complex 
scalar Higgs$'$ field $\phi$ responsible for spontaneous symmetry breaking. We imagine that any additional particles, in particular a possible WIMP dark matter 
candidate, are heavy compared to the vector and Higgs$'$. This new sector is not charged under the SM and vice versa, and all interactions with the SM proceed 
through kinetic mixing of \us\ with the photon (we can neglect  mixing with the $Z$ boson for the processes under consideration). The Lagrangian
then takes the form,
\begin{equation}
{\cal L}=-\frac{1}{4} V_{\mu\nu}^2 -\frac{\kappa}{2}\, V_{\mu\nu} F^{\mu\nu} + |D_\mu \phi |^2 -V(\phi),
\end{equation}
where $F_{\mu\nu}$ is the photon field strength, and the covariant derivative is $D_\mu=\partial_\mu+i e' V_\mu$ with \us\ charge $e'$. The
Higgs$'$ potential is assumed to be of a form which spontaneously breaks the \us\ symmetry. For example, neglecting mixing with the
SM Higgs which will be unimportant for low energy colliders, we can take $V(\ph) = -\mu^2|\ph|^2+\lambda|\ph|^4$, so that the Higgs$'$ 
acquires a vev $\langle \ph \rangle =  v'/\sqrt{2}$ with $v'=\sqrt{\mu^2/\lambda}$. Expanding around this vacuum, $\ph= (v'+h')/\sqrt{2}$, the unitary-gauge Lagrangian
containing the physical  Higgs$'$ field $h'$ takes the form,
\begin{equation}
{\cal L}=-\frac{1}{4} V_{\mu\nu}^2+\frac{1}{2}m_V^2 V_\mu^2 +\frac{1}{2}(\partial_\mu h')^2-\frac{1}{2}m_{h'}^2 h'^2 +{\cal L}_{int},
\end{equation}
where $m_V = e'v'$ and $m_{h'}=\sqrt{2\lambda} v'$. For completeness, the interaction terms are 
\begin{equation}
{\cal L}_{int}=-\frac{\kappa}{2}\,V_{\mu\nu}F^{\mu\nu}+\frac{m_V^2}{v'} h' V_\mu^2 +\frac{m_V^2}{v'^2}\,h'^2 V_\mu^2 - \frac{m_{h'}^2 }{2 v'} h'^3 - \frac{m_{h'}^2}{8 v'^2} h'^4, 
\end{equation}
although it is primarily just the $h' VV$ coupling that will be important in what follows. Note that
we can remove the kinetic mixing by redefining the photon $A_\mu\rightarrow A_\mu-\kappa V_\mu$, so that the SM fermions pick up a small U(1)$_S$
charge $\sim \kappa e$. Alternatively we can treat the kinetic mixing term as an interaction as long as $\kappa$ is small.

\subsubsection*{2.1 Decays}
In this section we will collect various expressions for the decay widths of $V_\mu$ and $h'$. As stated above, we assume that 
any other states coupled to $V_\mu$ and $h'$ are heavy ({\em e.g.} dark matter), so that all final states consist of 
SM particles. Many of the expressions given below can be adapted from the existing literature on $Z$ and Higgs boson phenomenology
(see {\em e.g.}~\cite{ew}).

\subsubsection*{2.1.1 $\Gamma_V$}

We begin with the decays of the vector $V_\mu$. Through mixing with the photon, $V_\mu$ will decay to SM leptons, with partial width
\begin{equation}
\Gamma_{ V \rightarrow \overline{l}l }=\frac{1}{3} \alpha \kappa^2 m_V \sqrt{1-\frac{4 m_l^2}{m_V^2}}
\left(1+\frac{2 m_l^2}{m_V^2}\right).
\end{equation}
Also, for masses $m_V> 2 m_\pi$, $V_\mu$ will decay to hadrons. Since $m_V$ may overlap with hadronic resonances, we 
will use the fact that the total decay width to hadrons can be directly related to the
 cross section $\sigma_{e^+ e^- \rightarrow {\rm hadrons}}$,
\begin{equation}
\Gamma_{ V \rightarrow {\rm hadrons} }=\frac{1}{3} \alpha \kappa^2 m_V \sqrt{1-\frac{4 m_\mu^2}{m_V^2}} \left(1+\frac{2 m_\mu^2}{m_V^2}\right) R(s= m_V^2),
\end{equation}
where as usual $ R= \sigma_{e^+ e^- \rightarrow {\rm hadrons}}/ \sigma_{e^+ e^- \rightarrow \mu^+ \mu^-}$. In the compilation of the hadronic cross section,
 the lowest data point is at $\sqrt{s}=0.36$ GeV \cite{pdg,hadron},  well above the pion threshold. Therefore in the intermediate range above the threshold we use the cross section for $e^+e^-\rightarrow \pi^+\pi^-$ \cite{hadron,pion}.

\begin{figure}
\centerline{
\includegraphics[width=1.05\textwidth]{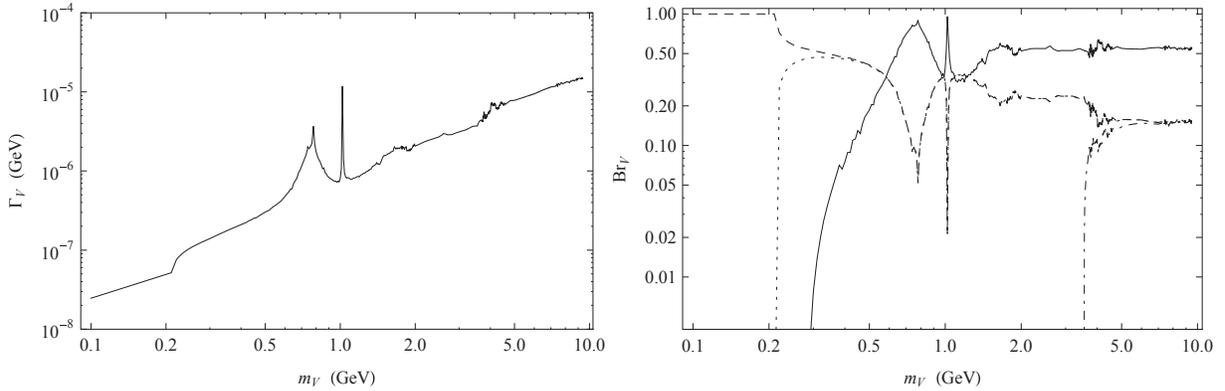}}
\caption{Total width $\Gamma_{V}$ (GeV) and branching ratios for $V$: $V\rightarrow  e^+ e^-$ (dashed), $V\rightarrow \mu^+ \mu^-$ (dotted), and $V\rightarrow  \tau^+ \tau^-$ (dot-dashed), and $V\rightarrow {\rm hadrons} $ (solid) for the choice of $\kappa=10^{-2}$ and $\alpha'=\alpha$. }
\label{figVwidth}
\end{figure}

We show in Fig.~\ref{figVwidth} the total $V$ decay width and branching ratios for $\kappa=10^{-2}$.
We see that for most values of $m_V$, the vector will have a significant branching to leptons (unless $m_V$ happens to coincide with a hadronic resonance). Note also that it is possible for the vector to decay to neutrinos due to kinetic mixing with the $Z$-boson, but this will be suppressed by a factor $m_V^4/m_Z^4 \sim 10^{-8}$ for a GeV-scale vector and can safely be neglected. 

\subsubsection*{2.1.2 $\Gamma_{h'}$}

Next we consider the decays of the Higgs$'$. The decay characteristics of the $h'$ depend on whether it is heavier or lighter than the vector. Let us first consider $m_{h'}>2m_V$, in which case the $h'$ decays predominantly to a pair of real vectors, with partial width
\begin{equation}
\Gamma_{h'\rightarrow VV}=\frac{\alpha' m^3_{h'}}{8 m_V^2}\sqrt{1-\frac{4 m^2_V}{m_{h'}^2}}
\left(1-\frac{4 m_V^2}{m_{h'}^2}+\frac{12 m_V^4}{m_{h'}^4}\right),
\label{hvv}
\end{equation}
where $\alpha'=e'^2/4\pi$. For the case $m_{h'} < m_V$, the Higgs$'$ can decay to leptons and hadrons via two off-shell vectors $V^*_\mu$ \cite{hvv}:
\begin{equation}
 \Gamma_{h'\rightarrow V^* V^*}=\frac{1}{\pi^2}
\int_0^{m_{h'}^2}\frac{dq_1^2 \, m_V \Gamma_V }{(q_1^2 - m_V^2)+m_V^2 \Gamma_V^2}
\int_0^{(m_{h'}-q_1)^2}\frac{dq_2^2 \, m_V \Gamma_V  }{(q_2^2-m_V^2)+m_V^2 \Gamma_V^2 }
\Gamma_0, 
\label{hvsvs}
\end{equation}
where
\begin{equation}
\Gamma_0=\frac{\alpha' m_{h'}^3}{8 m_V^2}\sqrt{\lambda\left(1,\frac{q_1^2}{m_{h'}^2},\frac{q_2^2}{m_{h'}^2}\right)}
\, \Bigg[ \lambda\left(1,\frac{q_1^2}{m_{h'}^2},\frac{q_2^2}{m_{h'}^2}\right)+\frac{12 q_1^2 q_2^2}{m_{h'}^4} \Bigg],
\end{equation}
with $\lambda(A,B,C)=A^2+B^2+C^2-2AB-2AC-2BC$. In fact Eq.~(\ref{hvsvs}) can also be used to calculate the the decay to one real and one virtual vector $h'\rightarrow V V^*$ for the regime $m_V<m_{h'} <2m_V$ as well as two real vectors (Eq. (\ref{hvv})) with the replacement of the Breit-Wigner peak by a delta-function.

\begin{figure}
\centerline{
\includegraphics[width=1.05\textwidth]{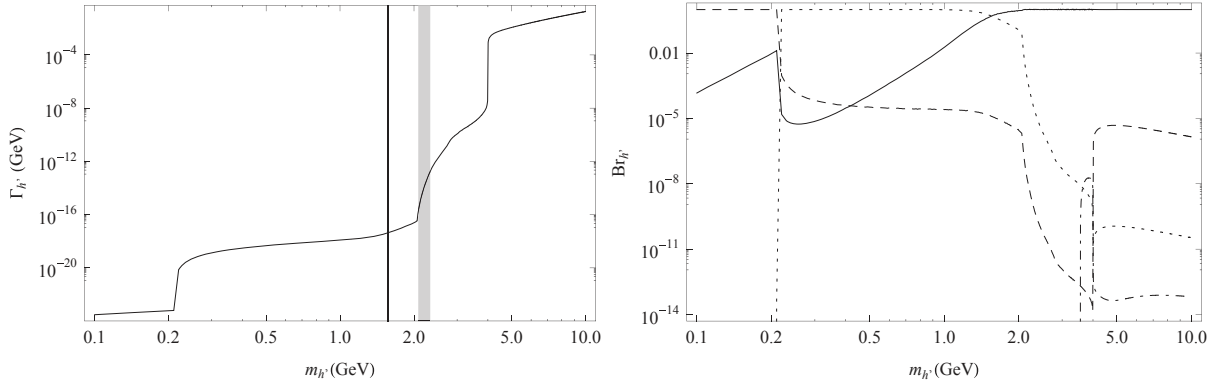}}
\caption{Total width $\Gamma_{h'}$ (GeV) and branching ratios for $h'$ for the case $m_V=2$ GeV: $h'\rightarrow  e^+ e^-$ (dashed), $h'\rightarrow \mu^+ \mu^-$ (dotted), $h'\rightarrow  \tau^+ \tau^-$ (dot-dashed), and $h'\rightarrow V V $ (solid) for the choice of $\kappa=10^{-2}$ and $\alpha'=\alpha$.
On the left panel, the vertical line delineates the boundary separating two- and four-particle final states, while the grey band defines the mass range
where the decay distance varies from 1\,mm to 1\,m leading to displaced vertices in the detector.
 }
\label{fighwidth}
\end{figure}

If the Higgs$'$ is light then loop induced decays become important. For example, the Higgs$'$ can decay into a pair of leptons through a triangle graph,
\begin{equation}
\Gamma_{h' \rightarrow \overline{f}f}= \frac{\alpha' \alpha^2 \kappa^4 m_{h'}}{2 \pi^2}\frac{m_f^2}{m_V^2}
\left(1-\frac{4 m_f^2}{m_{h'}^2}\right)^{3/2} |I(m_{h'},m_V,m_f)|^2,
\label{loop}
\end{equation}
where the form factor $I$ is
\begin{eqnarray}
I(m_{h'},m_V,m_f)&=&m_V^2\int_0^1 dx \int_0^{1-x} dy \frac{2-(x+y)}{(x+y)m_V^2 +[1-(x+y)]^2m_f^2- x y m_{h'}^2}. \nonumber \\ & &
\label{formfactor}
\end{eqnarray}
There are also loop-induced decays to hadrons for $m_{h'}>2 m_\pi$, which can compete with the leptonic modes (\ref{loop}). 
We have made a simple estimate for when these effects become important by calculating the decay $h'\rightarrow \pi^+ \pi^-$ using a Vector Meson Dominance model. We find that the two-pion channel begins to dominate  the dilepton channel (\ref{loop}) when $m_{h'} \gtrsim m_\rho$. 
However, as we discuss next, a detailed knowledge of hadronic decay mode properties will not be critical because if $h'$ is light it is
also long-lived and will manifest as missing energy in the detector.

The decay properties of the Higgs$'$ depend rather sensitively on the mass of the vector $m_V$. As an example, we plot the total width and branching fractions for the Higgs$'$ in Fig.~\ref{fighwidth} for a vector with  mass $m_V=2$ GeV. Heavy Higgs$'$ bosons will decay exclusively to vectors, which is expected as this is the only direct decay mode, with no $\kappa$ suppression. In turn these vectors will decay to leptons or hadrons.  However, light Higgs$'$ bosons with $m_{h'}< m_V$ will decay primarily through a loop-induced process, with a rate proportional to $\kappa^4 \times ({\rm loop~factor})^2 $. Although this is quite small, it still generically overwhelms the suppression from four-body phase space in the process $h'\rightarrow V^* V^* \rightarrow 4l$. Thus, 
when $h'$ is light, it is also extremely narrow and long-lived as indicated in Fig.~\ref{fighwidth}. As stated above, we have not taken into account the loop induced decays to hadrons for $m_{h'}> 2 m_\pi$, but we expect that the total width will not increase by more than an ${\cal O}(1)$ factor and therefore will not qualitatively change our results. The only exception to this is perhaps when $m_{h'}\sim m_V$ and the decay length is of the same order as the size of the detector, where a small increase in the width will allow $h'$ to decay inside the detector.

\subsection*{3. Collider signatures}

With an understanding of the decay properties of $V_\mu$ and $h'$, we move now to their production at electron-positron colliders. The $B$-factories BaBar and BELLE, with their mid-range center-of-mass energies $\sqrt{s}\sim 10$ GeV and large integrated luminosities $O(500 {\rm fb}^{-1})$, are very well suited to probing the minimal model discussed in the previous section, as well as extensions with other new particles at or below the GeV scale. 
However, because $V_\mu$ is quite narrow, single particle resonant production of vectors will be extremely unlikely unless the collider happens to be operating at $\sqrt{s}\sim m_V$. Furthermore, any anomalous contribution to the cross section of a generic process from an off-shell $V_\mu$ will be suppressed by $\kappa^4$ in the cross section. Therefore we will examine two-particle production mechanisms. 

One possibility is pair annihilation, $e^+e^- \rightarrow V \gamma$. This has the obvious advantage of being largely independent of details of the Higgs$'$ sector. Indeed this process was considered previously in  Ref.~\cite{BCD}, in relation to MeV-scale dark matter models \cite{mev,uboson1, uboson2}, and it was observed that the $B$-factories could detect vectors ($U$-bosons in their discussion) with mixings $\kappa \sim 10^{-3}$ to the SM. This study considered vectors somewhat lighter (tens of MeV) than those in which we are interested, but similar conclusions should apply for vectors in the GeV range. The on-shell vector will decay to a pair of leptons, leading to a signal of $l^+l^-\gamma$.  The SM background, although large for this process, is not a severe issue as the kinematics of the signal are quite distinct. The invariant mass of the lepton pair resides in a single bin due to the tiny width of the vector and can straightforwardly be distinguished from the background.

Besides  pair annihilation $e^+e^- \rightarrow V \gamma$, another particularly promising production channel is Higgs$'$-strahlung, $e^+ e^- \rightarrow V h'$, as it also is minimally suppressed by $\kappa^2$.
The total cross section for this process is 
\begin{eqnarray}
\sigma_{e^+ e^- \rightarrow V h'}&=&\frac{\pi \alpha \alpha' \kappa^2 }{3 s}\left(1-\frac{m_V^2}{s}\right)^{-2} \sqrt{\lambda\left(1,\frac{m^2_{h'}}{s},\frac{m^2_V}{s}\right)}
\Bigg[ \lambda\left(1,\frac{m^2_{h'}}{s},\frac{m^2_V}{s}\right)+ \frac{ 12 m_V^2}{s}\Bigg] \nonumber \\
&  \approx & 20 \, {\rm fb}  \times \left(\frac{\alpha'}{\alpha}\right) \left(\frac{\kappa^2}{10^{-4}} \right) \fr{(10~{\rm GeV})^2}{s},
\end{eqnarray}
where in the last line we have assumed the scaling regime, $m_{h'}+m_V < \sqrt{s}$ GeV, which is appropriate for WIMP models with a GeV-scale mediator. We can see that the cross section is quite sizable for reasonable values of the kinetic mixing parameter $\kappa$, with production of up to ten thousand $Vh'$ pairs given $500\, {\rm fb}^{-1}$ of data. Note that in our subsequent numerical calculations we will use the full expression for the cross section to account for 
the possibility of heavier vector and Higgs$'$ bosons. Note also that there is the chance of an $s$-channel enhancement in the cross section 
if $m_V \sim \sqrt{s}$.

Depending on the relative mass of $V$ and $h'$, the Higgs$'$-strahlung process will generate qualitatively distinct signals, as can be inferred from the branching fractions depicted in Figs.~\ref{figVwidth} and \ref{fighwidth}. If $h'$ is heavy, $m_{h'}>m_V$, the signal will consist of six leptons in the final state.  On the other hand, for $m_{h'}<m_V$, the Higgs$'$ will be long lived and escape the detector, resulting in two leptons plus missing energy.
As we will discuss next, although the total SM background cross sections can be quite large, the signal is distinguished by the kinematic properties of the final-state leptons 
as compared to the SM background, and we expect this kinematic information will be sufficient to extract evidence of Higgs$'$-strahlung, as it was for 
pair annihilation to $V\gamma$ \cite{BCD}. Moreover, we note that the Higgs$'$-strahlung process $e^+ e^- \rightarrow V h'\rightarrow 6l$ has (at least before applying any cuts)
 a significantly larger signal-to-background than the $V\gamma$ mode, $e^+ e^- \rightarrow V \gamma\rightarrow 2l\gamma$, as the background is $O(\alpha^6)$ for 
 the former, while it arises at $O(\alpha^3)$ for the latter.
However, we emphasize that we have not performed a proper calculation of the SM background with cuts, as our primary aim in this work is to emphasize the possibility of a large signal, which we hope will stimulate the experimental collaborations to perform more detailed analyses.

\subsubsection*{3.1 Signal and background} 

There are two cases to consider: either the Higgs$'$ is heavier than the vector, or vice-versa. These cases lead to  different experimental signatures. 

\subsubsection*{3.1.1 $m_{h'} >  m_V$}
First, for the case $m_{h'} \ge m_V$ the Higgs$'$ decays almost exclusively to two $V$'s, as illustrated in Fig.~\ref{fighwidth}. Since the vectors have a sizable branching fraction to leptons, this will lead to a signature of six leptons. For $m_{h'}>2m_V$ three pairs of leptons will have an invariant mass peaked very narrowly around the mass of the vector, 
$m_{l^+l^-} \simeq m_V$, reflecting an underlying decay $h'\to VV$ to on-shell vectors. For $2m_V>m_{h'}>m_V$,
the underlying decay is $h'\to VV^*$ and therefore two lepton pairs will have $m_{l^+l^-} \simeq m_V$. In addition
the four leptons will have a combined invariant mass of $m_{2(l^+l^-)} \simeq m_{h'}$. 

There are a few kinematic regimes to consider. For instance, if the vector is light,  $m_V < 2 m_\mu$, it can only decay to an $e^+ e^-$ pair, and thus every final state will consist of 3 $e^+ e^-$ pairs. The QED background for this process can be estimated using the equivalent photon approximation \cite{2photon},
\begin{eqnarray} 
\sigma_{e^+e^-\rightarrow 3(e^+e^-)}
&\approx & \frac{\alpha^6}{\pi^3 m^2}\left(\log{\frac{s}{m^2}} \right)^4 \\
& \approx &  10^9 ~{\rm fb},
\end{eqnarray}
leading to a huge background. However, in most events these are peripheral collisions and the two initial electrons will continue down the beam pipe, 
with $\theta \lesssim m / \sqrt{s}$. The signal events by contrast will have large transverse momentum, $p_T \sim$ few  GeV.  Thus a small 
angle cut should significantly reduce the background. Perhaps more importantly, the initial electron-positron pair will have an 
invariant masses $m_{e^+e^-}\sim \sqrt{s}$, compared to signal pairs with invariant masses $m_V$. Thus a suitable invariant mass cut 
 should allow the signal to be uncovered.

If $m_V>2m_\mu$, the vector can also decay to a $\mu^+ \mu^-$ pair, leading to a significant number of 6$\mu$ final states. The QED background in this case should be negligible as it cannot proceed via the two photon mechanism and the cross section will decrease as $1/s$. There will be a large number of fake events with $2e 6\mu$ arising from a two-photon process that can easily be distinguished since the $e^+e^-$ pair will be lost down the beam pipe leading to missing energy. Therefore, if the kinematic relation is right and triple muon pairs can be produced via Higgs$'$-strahlung, these events could be very efficiently used as a `discovery mode' for \us.  Above the pion threshold, the overall branching of $V$ to leptons will of course decrease, but 
there can still be a large number of multilepton events, especially for $\kappa\sim 10^{-2}$.

\subsubsection*{3.1.2 $m_{h'} <  m_V$}
Next, if $m_{h'}< m_V$, the Higgs$'$ is extremely narrow, with $\Gamma_{h'}\sim \kappa^4 \times ({\rm loop~factor})^2$, and thus 
very long-lived. For $m_V \gg  m_{h'} \gg 2 m_f$, the form factor (\ref{formfactor}) becomes $I\rightarrow 3/2$ and the lifetime of the $h'$ is
\begin{equation}
\tau_{h'} \sim 6 \times 10^{-9} \, {\rm s } \times \left( \frac{\alpha'}{\alpha}  \right)
 \left( \frac{\kappa}{10^{-2}}  \right)^{-4}
 \left( \frac{m_{h'}}{{\rm GeV}}  \right)
\left( \frac{m_V}{2 m_f}  \right)^2.
\end{equation}
Thus, $h'$ will be relativistic with a decay length on the order of ten meters or more, leading to a signal of two leptons plus missing energy. 
A suitable invariant mass cut on the lepton pair  should be able to reduce the SM background significantly. In passing, we note that the strong 
dependence of $\Gamma_{h'}$ on $\kappa$ means that it cannot be too small, as $\tau_{h'}$ can easily become comparable to 1 second 
for $\kappa \la 10^{-4}$. Such long lifetimes are excluded by $^4$He overproduction during Big Bang Nucleosynthesis. 

Although rather unlikely, in the range $m_{h'}\sim m_V$ it may be possible for the Higgs$'$ to decay in the detector to a $\mu^+\mu^-$ pair with a displaced vertex. This would mean a four muon signal of the underlying Higgs$'$-strahlung event. The QED background should be negligible as again there is no 
two-photon mechanism at work. 

Of course there could also be a $e^+e^-\mu^+\mu^-$ final state, with a large background from QED:
\begin{eqnarray} 
\sigma_{e^+e^-\rightarrow e^+e^-\mu^+\mu^-}
&\approx & \frac{\alpha^4}{ \pi m_\mu^2 }\left(\log{\frac{s}{m^2}} \right)^2 \log{\frac{s}{m_\mu^2}} \\
& \approx &  10^8 \, {\rm fb}.
\end{eqnarray}
As discussed previously, the electons in the two-photon process move primarily in the direction of the beam pipe. Small angle and invariant mass cuts should help to reduce the background. Moreover, the signal $h'$ decay events will have a significant displacement inside the detector, making 
 background rejection somewhat easier. 

It is important to point out that the possibility of observing two or four lepton events for the case $m_{h'} \lesssim m_V$ requires a more detailed analysis of the loop induced decays of $h'$ to hadrons. These decays increase the total width and will cause $h'$ to decay more readily within the 
detector but at the same time reduce the branching of $h'$ to leptons. If however the travel distance of $h'$  exceeds the size of the 
detector, the existence of an additional hadronic branching fraction will not affect the number 
of leptonic decays of $h'$ within the detector.

\begin{figure}
\centerline{\includegraphics[width=1.05\textwidth]{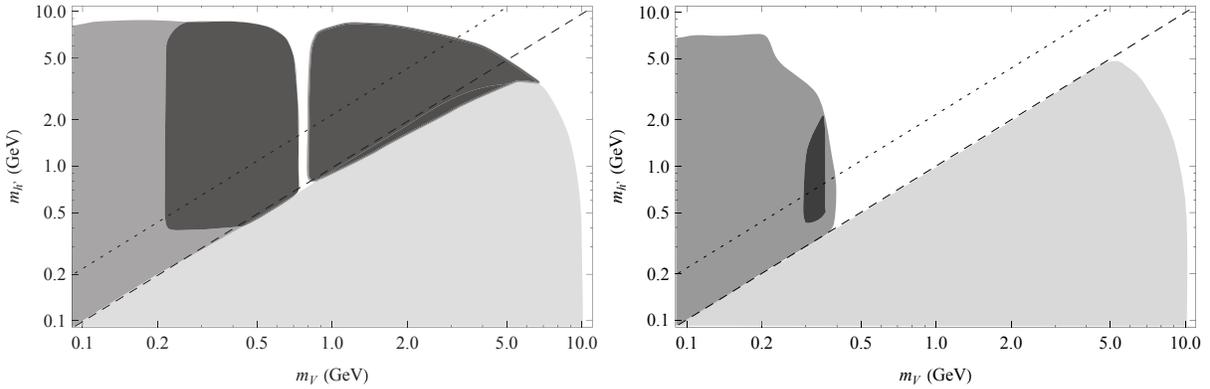}}
\caption{Sensitivity of the $B$-factory experiments for $\kappa = 10^{-2}$ (left) and $\kappa = 10^{-3}$ (right), with $\alpha'=\alpha$. The discovery modes 
are $l^+l^-$ plus missing energy (light), $6e$ (medium), and $6\mu$ (dark).
In the region $m_{h'} > 2 m_V \, (m_{h'} > m_V)$, above the dotted (dashed) line the signal contains $3\,(2)$ leptons pairs with 
invariant mass $m_{l^+l^-}\simeq m_V$.  }
\label{sens}
\end{figure} 

\subsubsection*{3.2 Results}
Our main results are presented in Fig.~\ref{sens}, which shows the maximum sensitivity of the $B$-factories to the Higgs$'$-strahlung channel, assuming 500 fb$^{-1}$ of integrated luminosity. These sensitivities assume a minimum of 10 signal events. We emphasize that
the SM backgrounds and various detector efficiencies were not taken into account in calculating these sensitivities, so that Fig.~\ref{sens} represents a ``best-case'' scenario with perfect efficiency and  in which the SM background can be completely eliminated by appropriate cuts without affecting the signal. Based on the discussion above, such cuts may indeed be possible, especially for the multilepton signal when $m_{h'}>m_V$.

Given these caveats, we find that BaBar and BELLE should have sufficient sensitivity to probe nearly all of the kinematically accessible $(m_{h'},m_V)$ parameter space for $\kappa\sim 10^{-2}$. In fact,  over a large part of the parameter space the number of signal events is in the thousands. For $\kappa\sim 10^{-3}$, the multilepton signal is useful for light vectors $m_V <500$ MeV, while the dilepton plus missing energy signal can be utilized for a light $h'$. 
An additional gain in experimental sensitivity could be achieved with the use of  
invariant mass information for the signal events, as discussed in the previous section. Finally, since the total 
number of the Higgs$'$-strahlung events scales as $\kappa^2\times$(integrated luminosity), the proposed
super-$B$ factories could reach the level of sensitivity $\kappa \sim O(10^{-4})$.

\subsection*{4. Discussion}

In summary, we have investigated the collider signatures of a secluded \us\ sector at the $B$-factories BaBar and BELLE, and find that these experiments should have an intrinsic
sensitivity to kinetic mixing angles of $\sim 10^{-3}$. We focused on the Higgs$'$-strahlung process, which has the advantage of leading to novel and thus
easily identifiable multi-lepton final states. Similar sensitivities will apply for pair-annihilation to $V\gamma$ as investigated in \cite{BCD}. We reiterate that 
our quoted sensitivities do not take into account SM backgrounds and cuts, as well as other issues such as detector efficiency which will necessarily 
reduce the total number of signal events. However, the number of signal events is at least of ${\cal O}(10^{3})$ over most of the relevant parameter 
space in Fig.~\ref{sens} for the case $\alpha'=\alpha$. Moreover, although tenuous at this point, if present hints toward 
WIMP dark matter at around 1~TeV were to solidify, then this would point to larger values for $\alpha'$ than are considered here (see {\em e.g.} \cite{PR}), allowing for
as much as a five-fold increase in the number of the Higgs$'$-strahlung events at fixed $\kappa$.  Based on these considerations, a full experimental 
analysis appears well warranted particularly given the degree of interest in secluded \us\ models.

\subsubsection*{4.1 Further signatures of a secluded sector}

We will finish by commenting on a number of other issues pertinent to searches for a \us\ sector.

\begin{itemize}

\item {\em Leptons vs pions.} Although we have concentrated on the leptonic signal of \us\ as perhaps the cleanest way of identifying Higgs$'$-strahlung events at $B$-factories, in certain kinematic domains the dominant decay modes will contain multi-pion final states, {\em e.g. } $3\pi^-3\pi^+$. Clearly, the pattern of invariant mass for multi-pions 
originating from $V$ or $h'$ decays will be exactly the same as for leptons. If such events are identified, and both muon and pion final states allow a reconstruction
from $V$ or $h'$ of a specific mass, the relative occurrence of lepton vs pion pairs will be regulated by $R(s= m_V^2)$. This is essentially independent of $\kappa$ 
and/or the structure of the Higgs$'$ sector, so that the pion vs lepton yield can be used as an additional cross-check of candidate \us\ events.

\item {\em Lepton jets.}
We have only considered the minimal model of an abelian U(1)$_S$ with a single complex scalar Higgs$'$ field. Already within this simple model it is 
quite common to have multi-lepton final states due to the Higgs$'$ decay $h'\rightarrow VV\rightarrow 4l$. These lepton pairs will have 
an angular separation $\theta \sim  m_{h'}/p_T$, and thus for a light Higgs$'$, $m_{h'}<$ GeV, may appear in a highly collimated ``lepton jet'' \cite{AW}. 
Note that this also requires $V$ to be light so that $h'$ decays predominantly to $VV$. Another plausible situation is a 
heavy Higgs$'$ bosons, $m_{h'} \sim$ GeV $> m_V$, in which case the lepton pairs will have a larger separation.

\item {\em  More complex secluded sectors and multi-lepton events.}
With more structure in the secluded sector, such as a non-abelian gauge group or multiple Higgs$'$ fields, the signals can be even more exotic. One 
rather generic prediction of additional light particles is the presence of complicated cascade decays, leading to multiple leptons 
(and hadrons if kinematically allowed). This has been appreciated in the ``hidden valley'' scenario \cite{hv}, and  
emphasized in \cite{AW} in the context of WIMP models with U(1)$_S$ mediators. As a simple example, suppose there is an 
additional scalar $S$ in the secluded sector, perhaps arising from a more complex Higgs$'$ sector,  that is heavier 
than $h'$. Production of $S$ may then produce complex decay patterns, {\em e.g.} $S\rightarrow 2 h'\rightarrow 4V \rightarrow 8l$ etc. 
Similar phenomena may occur with additional gauge bosons from a non-abelian hidden sector \cite{AW, nonabelian}. It would of course be 
worthwhile to consider the collider signatures of more complex secluded sectors at medium-energy $e^+e^-$ colliders such as the $B$-factories. 

\item {\em Higgs$'$-strahlung events at hadron colliders.} 
The Higgs$'$-strahlung process may also be of considerable interest for hadron collider signatures of secluded \us\ sectors. The advantage of the
$B$-factories comes from the combination of an optimal center-of-mass energy for a GeV-scale \us\ with large integrated luminosities, as well as, of course, much better control over the background. Hadron colliders
such as the Tevatron and the LHC have to make up for the production rate falling as $1/s$ and generally lower luminosity. However, production processes
such as $q\overline{q}\rightarrow Vh'$, and those involving gluons in the initial state, can take advantage of the rapid rise of parton distribution functions at
small $x$. Initial estimates suggest that this enhancement may well be sufficient to make an analysis of Higgs$'$-strahlung at the Tevatron and
the LHC of further interest \cite{bpr2}. 

\item {\em Long-lived Higgs$'$,  cosmic rays, and neutrino detectors.} The longevity of $h'$ for $m_{h'} < m_V$ can lead to a long decay distance. While 
such a light Higgs$'$ will simply appear as missing energy at colliders, it may well produce easily identifiable multi-lepton signals in neutrino detectors such as
super-K, IceCube, Antares, and possibly MiniBoone for decay distances of a few hundred meters. For example, cosmic rays could induce 
Higgs$'$-strahlung events in the upper atmosphere which, although suppressed by $\kappa^2$, are capable of producing two- or four-muon events from $h'$ decay 
inside IceCube, creating an unmistakable signature. It is important to note that there is a well-defined window for decays, as lifetimes longer than about 1 second will
be ruled out by BBN bounds. Thus the regime $m_{h'}<m_V$, which could be particularly relevant for supersymmetric extensions of \us, may be most usefully probed using 
large-scale neutrino detectors of various forms.

\item {\em Supersymmetric extensions of \us.} A supersymmetric \us\ sector may allow the most natural explanation for the $m_V \ll m_Z$ hierarchy \cite{HZ,nonabelian}. 
This necessarily introduces more Higgs$'$ bosons as well as extra Higgsino$'$ and gaugino$'$ states. The detailed phenomenology is then quite complex, {\em e.g.} production of extra heavy Higgs$'$ states may lead to more complex multi-lepton final states, while the new gaugino-Higgsino sector opens the possibility for less visible 
Higgs$'$ decays. We will note here only the generic point that kinetic mixing in the form $\int d^2\theta\, W_S W_Y$, which as noted in \cite{nonabelian} leads 
to a generic form of SUSY mediation, will also imply via the associated $D$-terms an intrinsic quartic coupling between the 
SM Higgs and Higgs$'$ sectors. Although suppressed by $\ka$, this will likely imply non-negligible constraints from flavor physics  \cite{PRV}.

\item{\em Resonant $s$-channel production of $V$.} Finally, we remark that although the main emphasis of this work was on the possibility of a \us\ search 
at  $e^+e^-$ machines of medium energy, the higher end of the kinematic range that we consider -- $m_V\sim$ few GeV and large values of 
$\kappa \sim 10^{-2}$ -- corresponds to resonances as broad as a few keV, and therefore allows for a search using a high-energy resolution scan of the relevant 
energy domain \cite{tests}.

\end{itemize}

\subsubsection*{4.2. Concluding remarks}

The high level of attention focussed recently on secluded \us\ scenarios has been driven primarily by their interesting phenomenology
in the context of dark matter \cite{PRV,AFSW,PR,HZ}, but such  models are well-motivated in their own right: kinetic mixing via the 
vector portal \cite{holdom} provides one of the few renormalizable probes  of a hidden sector.  
As we have shown, $B$-factories allow for a decisive test for the presence of a GeV-scale secluded \us, probing kinetic mixing with the photon down 
to the level of $O(10^{-3})$. In this way $B$-factories can, within existing data sets, significantly improve upon other precision QED and flavor physics tests of 
such models. The striking signature of multi-lepton events provides additional motivation for  the BaBar and BELLE collaborations to perform
dedicated analyses.

\subsection*{Acknowledgements}

We thank G. Franzoni, R. Kowalewski, M. Roney, and M. Trott  for helpful discussions.
The work of A.R. and M.P. is supported in part by NSERC, Canada, and research at the Perimeter Institute
is supported in part by the Government of Canada through NSERC and by the Province of Ontario through MEDT.


\begin{thebibliography}{99}


\bi{holdom}
  B.~Holdom,
  Phys.\ Lett.\  B {\bf 166}, 196 (1986).

\bi{review}
{\it see {\em e.g.}} G.~Jungman, M.~Kamionkowski and K.~Griest,
  Phys.\ Rept.\  {\bf 267}, 195 (1996)
  [arXiv:hep-ph/9506380];
G.~Bertone, D.~Hooper and J.~Silk,
  Phys.\ Rept.\  {\bf 405}, 279 (2005)
  [arXiv:hep-ph/0404175].
  
\bi{PRV}  M.~Pospelov, A.~Ritz and M.~B.~Voloshin,
  Phys.\ Lett.\  B {\bf 662}, 53 (2008)
  [arXiv:0711.4866 [hep-ph]].

\bi{AFSW}
  N.~Arkani-Hamed, D.~P.~Finkbeiner, T.~Slatyer and N.~Weiner,
  arXiv:0810.0713 [hep-ph].

\bi{PR}
  M.~Pospelov and A.~Ritz,
  Phys.\ Lett.\  B {\bf 671}, 391 (2009)
  [arXiv:0810.1502 [hep-ph]].



\bi{cholis} 
  I.~Cholis, L.~Goodenough and N.~Weiner,
  arXiv:0802.2922 [astro-ph].


\bi{pamela}
  O.~Adriani {\it et al.},
  arXiv:0810.4995 [astro-ph].
  

\bi{atic}
  J.~Chang {\it et al.},
  Nature {\bf 456}, 362 (2008).
  
  
\bi{tests}
  M.~Pospelov,
  arXiv:0811.1030 [hep-ph].
  


\bi{g-2}
   F.~Jegerlehner and A.~Nyffeler,
  arXiv:0902.3360 [hep-ph].

\bi{mev}
  C.~Boehm and P.~Fayet,
  Nucl.\ Phys.\  B {\bf 683}, 219 (2004)
  [arXiv:hep-ph/0305261].


\bi{uboson1}
  P.~Fayet,
  Phys.\ Rev.\  D {\bf 74}, 054034 (2006)
  [arXiv:hep-ph/0607318].


\bi{uboson2}
  P.~Fayet,
  Phys.\ Rev.\  D {\bf 75}, 115017 (2007)
  [arXiv:hep-ph/0702176].
  
\bi{BCD}
  N.~Borodatchenkova, D.~Choudhury and M.~Drees,
  Phys.\ Rev.\ Lett.\  {\bf 96}, 141802 (2006)
  [arXiv:hep-ph/0510147].

\bibitem{AW}
  N.~Arkani-Hamed and N.~Weiner,
  JHEP {\bf 0812}, 104 (2008)
  [arXiv:0810.0714 [hep-ph]].

\bi{nonabelian}
  M.~Baumgart, C.~Cheung, J.~T.~Ruderman, L.~T.~Wang and I.~Yavin,
  arXiv:0901.0283 [hep-ph];  C.~Cheung, J.~T.~Ruderman, L.~T.~Wang and I.~Yavin,
  arXiv:0902.3246 [hep-ph].
 
 \bi{langacker}
  P.~Langacker,
  arXiv:0801.1345 [hep-ph].
  
  
    \bi{nw}
  A.~E.~Nelson and J.~Walsh,
  Phys.\ Rev.\  D {\bf 77}, 033001 (2008)
  [arXiv:0711.1363 [hep-ph]].
  
  \bi{prv2}
  M.~Pospelov, A.~Ritz and M.~B.~Voloshin,
  Phys.\ Rev.\  D {\bf 78}, 115012 (2008)
  [arXiv:0807.3279 [hep-ph]].
  
  \bi{jrr}
  J.~Jaeckel, J.~Redondo and A.~Ringwald,
  Phys.\ Rev.\ Lett.\  {\bf 101}, 131801 (2008)
  [arXiv:0804.4157 [astro-ph]]; 
  M.~Ahlers, J.~Jaeckel, J.~Redondo and A.~Ringwald,
  Phys.\ Rev.\  D {\bf 78}, 075005 (2008)
  [arXiv:0807.4143 [hep-ph]]; 
  J.~Redondo and M.~Postma,
  JCAP {\bf 0902}, 005 (2009)
  [arXiv:0811.0326 [hep-ph]].


\bi{ew}
  A.~Djouadi,
  Phys.\ Rept.\  {\bf 457}, 1 (2008)
  [arXiv:hep-ph/0503172].

\bi{pdg}
  C.~Amsler {\it et al.}  [Particle Data Group],
  Phys.\ Lett.\  B {\bf 667}, 1 (2008).

\bibitem{hadron}
  V.~V.~Ezhela, S.~B.~Lugovsky and O.~V.~Zenin,
  arXiv:hep-ph/0312114.

\bibitem{pion}
  M.~Davier, S.~Eidelman, A.~Hocker and Z.~Zhang,
  Eur.\ Phys.\ J.\  C {\bf 27}, 497 (2003)
  [arXiv:hep-ph/0208177].


\bi{hvv}
  R.~N.~Cahn,
  Rept.\ Prog.\ Phys.\  {\bf 52}, 389 (1989);
  A.~Grau, G.~Panchieri and R.~J.~N.~Phillips,
  Phys.\ Lett.\  B {\bf 251}, 293 (1990);
  B.~A.~Kniehl,
  Phys.\ Lett.\  B {\bf 244}, 537 (1990).


\bi{2photon}
  V.~M.~Budnev, I.~F.~Ginzburg, G.~V.~Meledin and V.~G.~Serbo,
  Phys.\ Rept.\  {\bf 15}, 181 (1974).
  
  \bi{hv}
  M.~J.~Strassler and K.~M.~Zurek,
  Phys.\ Lett.\  B {\bf 651}, 374 (2007)
  [arXiv:hep-ph/0604261].
  M.~J.~Strassler,
  arXiv:hep-ph/0607160.
  
  \bi{bpr2}
  B.~Batell, M.~Pospelov, and A.~Ritz, {\it work in progress}.

\bi{HZ}  D.~Hooper and K.~M.~Zurek,
  Phys.\ Rev.\  D {\bf 77}, 087302 (2008)
  [arXiv:0801.3686 [hep-ph]].
  
  



  \end{thebibliography}
\end{document}